
%
%
%
%
\input harvmac
%
%
%
%
\ifx\answ\bigans
\else
\output={
  \almostshipout{\leftline{\vbox{\pagebody\makefootline}}}\advancepageno
}
\fi
%
%
%

%
%

%
%
\def\UCSD#1#2{\noindent#1\hfill #2%
\supereject\global\hsize=\hsbody%
\footline={\hss\tenrm\folio\hss}}
%
%
\def\abstract#1{\centerline{\bf Abstract}\nobreak\medskip\nobreak\par #1}
%
%
%
%
\edef\tfontsize{ scaled\magstep3}
 \tfontsize  \tfontsize
 \tfontsize \font\titlei=cmmi10 \tfontsize
\font\titleis=cmmi7 \tfontsize \font\titleiss=cmmi5 \tfontsize
\font\titlesy=cmsy10 \tfontsize \font\titlesys=cmsy7 \tfontsize
\font\titlesyss=cmsy5 \tfontsize  \tfontsize
\skewchar\titlei='177 \skewchar\titleis='177 \skewchar\titleiss='177
\skewchar\titlesy='60 \skewchar\titlesys='60 \skewchar\titlesyss='60
%
%
%
%
%
\def\inv{^{\raise.15ex\hbox{${\scriptscriptstyle -}$}\kern-.05em 1}}
\def\lbar{{\lower.35ex\hbox{$\mathchar'26$}\mkern-10mu\lambda}} 

%
%
%
%
\def\slash#1{\rlap{$#1$}/} 
\def\dsl{\,\raise.15ex\hbox{/}\mkern-13.5mu D} 
\def\delsl{\raise.15ex\hbox{/}\kern-.57em\partial}
\def\Ksl{\hbox{/\kern-.6000em\rm K}}
\def\Asl{\hbox{/\kern-.6500em \rm A}}
\def\Dsl{\hbox{/\kern-.6000em\rm D}} 
\def\Qsl{\hbox{/\kern-.6000em\rm Q}}
\def\gradsl{\hbox{/\kern-.6500em$\nabla$}}
%
%
\def\lspace{\ifx\answ\bigans{}\else\qquad\fi}
\def\lbspace{\ifx\answ\bigans{}\else\hskip-.2in\fi} 
%
%
\def\boxeqn#1{\vcenter{\vbox{\hrule\hbox{\vrule\kern3pt\vbox{\kern3pt
        \hbox{${\displaystyle #1}$}\kern3pt}\kern3pt\vrule}\hrule}}}
%
%
\def\mbox#1#2{\vcenter{\hrule \hbox{\vrule height#2in
\kern#1in \vrule} \hrule}}
%
%
%
%
\def\CA{{\cal A}}  \def\CC{{\cal C}} \def\CD{{\cal D}}
   \def\CH{{\cal H}}
   \def\CL{{\cal L}}
\def\CM{{\cal M}}  \def\CO{{\cal O}} 
   
\def\CU{{\cal U}}   
 
%
%
%
%
%

%

\def\bar#1{\overline{#1}}

\def\darr#1{\raise1.5ex\hbox{$\leftrightarrow$}\mkern-16.5mu #1}

%
%
\def\half{{\textstyle{1\over2}}} 
\def\frac#1#2{{\textstyle{#1\over #2}}} 
%
%
%
%

\def\Tr{\mathop{\rm Tr}}

%
%
%
%

%
%
\def\ltap{\ \raise.3ex\hbox{$<$\kern-.75em\lower1ex\hbox{$\sim$}}\ }
\def\gtap{\ \raise.3ex\hbox{$>$\kern-.75em\lower1ex\hbox{$\sim$}}\ }
\def\gl{\ \raise.5ex\hbox{$>$}\kern-.8em\lower.5ex\hbox{$<$}\ }
\def\roughly#1{\raise.3ex\hbox{$#1$\kern-.75em\lower1ex\hbox{$\sim$}}}
%
%
        
\def\eg{\hbox{\it e.g.}}        
\def\etal{\hbox{\it et al.}}

\def\np#1#2#3{Nucl. Phys. B{#1} (#2) #3}
\def\pl#1#2#3{Phys. Lett. {#1}B (#2) #3}
\def\prl#1#2#3{Phys. Rev. Lett. {#1} (#2) #3}
\def\physrev#1#2#3{Phys. Rev. {#1} (#2) #3}

\relax
\ifx\epsfbox\notincluded\message{(NO epsf.tex, FIGURES WILL BE IGNORED)}
\def\insertfig#1#2{}
\else\message{(FIGURES WILL BE INCLUDED)}\def\insertfig#1#2{
\midinsert\centerline{\epsffile{#2}}\centerline{{#1}}\endinsert}\fi
\noblackbox

\def\a{\alpha}

\def\b{\beta}
\def\b{\beta}
\def\bpi{\beta^{(\pi)}}
\def\bk{\beta^{(K)}}

\def\c{\CC}
\vskip 1.in
\centerline{{\titlefont{Chiral Perturbation Theory Analysis of}}}
\medskip
\centerline{{\titlefont{the Baryon Magnetic Moments}}}
\vskip .2in
\centerline{Elizabeth Jenkins,${}^a$\footnote{${}^*$}{On leave from the
University of California at San Diego.} Michael Luke,${}^b$}
\smallskip
\centerline{Aneesh V.~Manohar,${}^{a\,*}$ and
Martin J. Savage${}^b$\footnote{$^{\dagger}$}{SSC Fellow}}
\medskip
\centerline{\sl a) CERN TH Division, CH-1211 Geneva 23, Switzerland}
\smallskip
\centerline{\sl b) Department of Physics, University of California at San
Diego,}\centerline{\sl 9500 Gilman Drive, La Jolla, CA 92093}
\vfill
\abstract{
Nonanalytic $m_q^{1/2}$ and $m_q\ln m_q$ chiral corrections to the
baryon magnetic moments are computed.  The calculation includes
contributions from both intermediate octet and decuplet baryon states.
Unlike the one-loop contributions to the baryon axial currents and
masses, the contribution from decuplet intermediate  states does not
partially cancel that from octet intermediate states. The fit to the
observed magnetic moments including $m_q^{1/2}$ corrections is found to
be much worse than the tree level $SU(3)$ fit if values for the
baryon-pion axial coupling constants obtained from a tree level
extraction are used. Using the axial coupling constant values extracted
at one loop results in a better  fit to the magnetic moments than the
tree level $SU(3)$ fit. There are three linear relations amongst the
magnetic moments when $m_q^{1/2}$ corrections are included, and one
relation including $m_q^{1/2}$, $m_q\ln m_q$ and $m_q$ corrections.
These relations are independent of the axial coupling constants of the
baryons and agree well with experiment.
}
\vfill
\UCSD{\vbox{
\hbox{CERN-TH.6735/92}\vskip-0.1truecm
\hbox{UCSD/PTH 92-34}\vskip-0.1truecm
\hbox{hep-ph/9212226}}}{December 1992}
\eject

The baryon magnetic moments were first predicted theoretically on the
basis of flavor $SU(3)$ symmetry \ref\coleglas{S. Coleman and S.L.
Glashow, \prl{6}{1961}{423}}. The Coleman-Glashow $SU(3)$ relations
yield the octet baryon magnetic moments in terms of two parameters.
These relations are easily derived from the Lagrangian
\eqn\cgmag{
\CL = {e \over {4 m_N}} \left( \mu_D \Tr \bar B_v \sigma_{\mu \nu}
F^{\mu \nu} \left\{ Q , B_v \right\} + \mu_F \Tr \bar B_v
\sigma_{\mu \nu} F^{\mu \nu} \left[ Q , B_v \right] \right),
}
where the parameters $\mu_{D,F}$ multiply the two independent $SU(3)$
invariants, $m_N$ is the nucleon mass, and
\eqn\qcharge{
Q = \pmatrix{\frac 2 3 &0 &0 \cr
              0 &-\frac 1 3 &0 \cr
              0 &0 &-\frac 1 3 \cr
},}
is the charge matrix for the three light quarks $u,\,d$ and $s$.
Calculation of the magnetic moments beyond tree approximation is
possible in chiral perturbation theory \ref\calpag{D.G. Caldi and H.
Pagels, \physrev{D10}{1974}{3739}}. The leading corrections which occur
at one loop have a non-analytic dependence on the light quark masses of
the form $m_q^{1/2}$ and $m_q \ln m_q$.  The $m_q^{1/2}$ and $m_q\ln
m_q$ terms for the baryon magnetic moments are calculable since they are
non-analytic in the quark masses \ref\lpagels{L-F. Li and H. Pagels,
\prl{26}{1971}{1204} \semi P. Langacker and H. Pagels,
\physrev{D10}{1974}{2904} \semi H. Pagels, Phys. Rep. 16 (1975)
219}\ref\gasleut{J. Gasser and H. Leutwyler, Phys. Rep. 87 (1982) 77}\
and therefore can not arise from terms in the chiral Lagrangian with
additional insertions of the quark mass matrix. The $m_q^{1/2}$
contribution \calpag\ref\gss{J. Gasser, M. Sainio and A. Svarc,
\np{307}{1988}{779}}\ and $m_q \ln m_q$ contribution \ref\krause{A.
Krause, Helv.~Phys.~Acta. 63 (1990) 3}\ from intermediate octet states
have been computed previously. Recent work on baryon chiral perturbation theory
\ref\jm{E. Jenkins and A.V. Manohar, \pl{255}{1991}{558}, \pl{259}{1991}{353}}%
\nref\j{E. Jenkins, Nucl. Phys. B368 (1992) 190, Nucl. Phys. B375 (1992) 561}%
\nref\jmtwo{E. Jenkins and A.V. Manohar, {\sl Baryon Chiral Perturbation
Theory}, in Proceedings of the Workshop on ``Effective Field Theories of the
Standard Model,'' ed. U. Meissner, World Scientific (1992) }%
\nref\bsone{M.N. Butler and M.J. Savage, UCSD Preprint UCSD/PTH 92-30
(1992) [{\tt hep-ph/9209204}]}--\ref\cb{T.D. Cohen and W. Broniowski,
\pl{292}{1992}{5}, and University of Maryland Preprints 92-193 and
92-225 (1992) [{\tt hep-ph/9208256-7}]}\ indicates that the spin-$3/2$
decuplet of baryons contributes significantly as an intermediate state
in one-loop diagrams with octet baryon initial and final states. In this
letter, we calculate the $m_q^{1/2}$ and $m_q\ln m_q$ contributions to
the baryon magnetic moments including both intermediate octet and
intermediate decuplet states.

The chiral Lagrangian for baryon fields depends on the pseudoscalar
pion octet
\eqn\pion{
\pi = {1 \over {\sqrt{2}}} \pmatrix{ {1\over\sqrt2}\pi^0 +
{1\over\sqrt6}\eta&
\pi^+ & K^+\cr
\pi^-& -{1\over\sqrt2}\pi^0 + {1\over\sqrt6}\eta&K^0\cr
K^- &\bar K^0 &- {2\over\sqrt6}\eta\cr
},
}
which couples to the baryon matter fields through the vector and
axial vector combinations
\eqn\av{
V^{\mu} = \frac 1 2 ( \xi \partial^{\mu} \xi^{\dagger}
+ \xi^\dagger \partial^{\mu} \xi ) , \qquad
A^{\mu} = \frac i 2 (\xi \partial^{\mu} \xi^{\dagger}
- \xi^\dagger \partial^{\mu} \xi ),
}
where
\eqn\sigmaxi{
\xi = e^{i\pi/f}, \quad \Sigma = \xi^2 = e^{2 i \pi/f},
}
and $f \approx 93$~MeV is the pion decay constant.
Under $SU(3)_L \times SU(3)_R$ chiral symmetry,
\eqn\tran{\eqalign{
&\Sigma \rightarrow L \Sigma R^\dagger, \qquad
\xi \rightarrow L \xi U^\dagger = U \xi R^\dagger, \cr
&B \rightarrow U B U^\dagger,\qquad
 T^{\mu}_{abc} \rightarrow U^d_a\, U^e_b\, U^f_c\,
T^{\mu}_{def}, \cr
}}
where $U$ is defined by the transformation of $\xi$, and $B$ and
$T^{\mu}$ denote the baryon octet and decuplet fields, respectively.  A
consistent chiral derivative expansion for baryon fields \jm\ can be
written in terms of velocity-dependent baryon fields,
\eqn\pbv{\eqalign{
 T^{\mu}_v (x) &= e^{i m_B \slash v \, v_{\mu} x^{\mu}}
T^{\mu}(x), \cr
B_v (x) &= e^{i m_B \slash v \, v_{\mu} x^{\mu} } B(x), \cr
}}
where $m_B$ is the $SU(3)$ invariant mass of the octet baryon
multiplet.  The lowest order chiral Lagrangian for octet and
decuplet baryons is
\eqn\vlag{\eqalign{
L^0_v &= i \Tr \bar B_v (v \cdot \CD) \ B_v
+ 2 D \Tr \bar B_v S_v^{\mu} \left\{ A_{\mu}, B_v \right\}
+ 2 F \Tr \bar B_v S_v^{\mu} \left[ A_{\mu}, B_v \right] \cr
&-i \, \bar T^{\mu}_v (v \cdot \CD) \ T_{v\,\mu} + \CC\,
\left(\bar T^{\mu}_v A_{\mu} B_v
+ \bar B_v A_{\mu} T_v^{\mu}\right)
+2 \CH \,\bar T^{\mu}_v S_{v \, \nu} A^{\nu} T_{v \, \mu}
\cr &+ \delta \,\bar T_v^{\mu} T_{v \, \mu}
+ {{f^2}\over 4} \Tr \partial_{\mu} \Sigma \partial^{\mu}
\Sigma^{\dagger} , \cr
}}
where the decuplet-octet mass difference $\delta= m_T - m_B$, and $m_T$
is the $SU(3)$ invariant mass of the baryon decuplet. The vector
combination of pion fields appears in Eq.~\vlag\ through the covariant
derivatives
\eqn\covd{\eqalign{
&\CD^{\nu} B = \partial^{\nu} B + [V^{\nu}, B] \cr
&\CD^{\nu} T^{\mu}_{abc} = \partial^{\nu}  T^{\mu}_{abc}
+ (V^{\nu})^d_a  T^{\mu}_{dbc} + (V^{\nu})^d_b
 T^{\mu}_{adc} + (V^{\nu})^d_c  T^{\mu}_{abd}. \cr
}}
The axial vector pion couplings are described by four coupling
constants $D, F, \CC$~and $\CH$.  The octet and decuplet
baryon propagators obtained from this Lagrangian are
$i/ (v \cdot k)$ and $i P_v^{\mu \nu}/ (v \cdot k - \delta )$, where
\eqn\p{
P_v^{\mu \nu} = \sum_i \CU_i^{\mu} \bar\CU_i^{\nu}
=\left( v^{\mu} v^{\nu} - g^{\mu \nu} \right) - \frac 4 3
S_v^{\mu} S_v^{\nu} .
}
is a polarization projector for the spin-$3/2$ Rarita-Schwinger
decuplet field.  The decuplet field satisfies the constraint
$v \cdot T =0$.

The calculation of the baryon magnetic moments involves
electromagnetic couplings.  Electromagnetism is incorporated into
Lagrangian~\vlag\ by making the following substitutions,
\eqn\emsub{\eqalign{
&V^{\mu} \rightarrow V^{\mu} + \frac 1 2 i e \CA^\mu
\left( \xi^\dagger Q \xi + \xi Q \xi^\dagger \right), \cr
&A^{\mu} \rightarrow A^{\mu} - \frac 1 2 e \CA^\mu
\left( \xi Q \xi^\dagger - \xi^\dagger Q \xi \right) , \cr
}}
and
\eqn\covsig{
\partial_\mu \Sigma \rightarrow \CD_\mu \Sigma =
\partial_\mu \Sigma + i e \CA_\mu [Q, \Sigma] \, ,
}
where $\CA_\mu$ is the photon field.
The octet baryon magnetic moment Lagrangian is given in Eq.~\cgmag;
the full chiral structure of the operator is given by the replacement
\eqn\repl{
Q\rightarrow \half( \xi Q \xi^\dagger + \xi^\dagger Q \xi).
}
The one loop corrections also involve the decuplet magnetic moment and
the decuplet-octet transition magnetic moment. There is only one $SU(3)$
invariant in the tensor product ${\bf \bar {10} \otimes 10 \otimes 8}$,
which can be chosen to be proportional to the charge, so
the decuplet magnetic moment operator can be written in the form
\eqn\decmag{ \CL=-i {e\over
m_N} \mu_C\ q_i\ \bar T^\mu_{v\,i}\ T^\nu_{v\,i} F_{\mu\nu}, }
where
$q_i$ is the charge of the $\imath^{\rm th}$ element of the decuplet,
and the operator is normalized so that the magnetic moment of the
$\imath^{\rm th}$ state is $q_i \mu_C$ nuclear magnetons. The measured
value of the $\Omega^-$ magnetic moment \ref\diehl{H.T. Diehl, \etal,
\prl{67}{1991}{804}}\ determines $\mu_C=1.94\pm0.22$. The octet-decuplet
transition magnetic moment operator has the form \ref\bss{M.N. Butler,
M.J. Savage, and R.P. Springer, Strong and Electromagnetic Decays of the Baryon
Decuplet, UCSD Preprint UCSD/PTH 92-37 (1992) [{\tt hep-ph/9211247}]}
\eqn\transmag{ \CL = i
{e\over 2 m_N}\mu_T\ F_{\mu \nu}\left( \epsilon_{ijk}\
Q^i_l\ \bar B^j_{v\,m}\ S^\mu_v\ T^{\nu klm}_v
+\epsilon^{ijk}\ Q^l_i\ \bar T^{\mu}_{v\,klm}\ S^\nu_v\ B^m_{v\,j}  \right),
}
where ${i,j,k,l,m}$ are $SU(3)$ flavor indices. The octet-decuplet transition
electric quadrupole moment operator has an additional derivative, and
therefore is higher order in the chiral expansion. The Lagrangian
Eq.~\transmag\ implies that the $\Delta \rightarrow N\gamma$ helicity
amplitudes $A_{3/2}$ and $A_{1/2}$ are in the ratio $\sqrt 3 : 1$, in
good agreement with experiment \ref\blp{M.A.B.~B\'eg, B.W.~Lee and
A.~Pais, \prl{13}{1964}{514}\semi C.~Becchi and G.~Morpurgo,
Phys.~Lett.~17 (1965) 352}\ref\hamp{Review of Particle Properties,
\physrev{D45}{1992}{VIII.16}}. The measured values of the helicity
amplitudes determine $\mu_T=-7.7\pm0.5$.

In addition to the electromagnetic couplings, the calculation
of the magnetic moments to nonleading order requires the
introduction of $SU(3)$ breaking through the quark mass matrix
$\CM = {\rm diag}\,(m_u, m_d, m_s)$.  At leading order in the quark mass
expansion, the  pseudo-Goldstone bosons acquire non-vanishing masses
and the $SU(3)$ baryon multiplets are no longer degenerate.

The calculation of the baryon magnetic moments presented here
includes the tree-level Coleman-Glashow formul\ae\ and the leading
non-analytic correction arising from the diagrams displayed in
\fig\figone{Diagrams which produce the non-analytic $m_q^{1/2}$
contribution to the baryon magnetic moments.  Dashed lines denote
pions; single and double solid lines denote octet and decuplet
baryons, respectively.}\ and \fig\figtwo{Diagrams which produce the
non-analytic $m_q\ln m_q$ contribution to the baryon magnetic moments
(excluding the wavefunction renormalization graphs). Dashed lines denote
pions; single and double solid lines denote octet and decuplet baryons,
respectively. The photon vertices in (a) and (b) are from the octet
baryon moment, in (c) from the decuplet magnetic moment, and in (d) and
(e) from the decuplet-octet transition magnetic moment. }.\foot{There
are also graphs which involve the $\bar B B \pi \CA_\mu$ vertex arising
from the $Q$ terms in Eq.~\emsub. These graphs do not contribute to the
magnetic moments.} The magnetic moments (in units of nuclear magnetons)
can be written in the form
\eqn\magb{\eqalign{
\mu_{i} = { \a}_i + \sum_{X=\pi,K}
\beta_i^{(X)} {M_{X} m_N\over  8 \pi f^2}+
\sum_{X=\pi,K} F\left(M_X,\delta,\mu\right)
\beta_i^{'(X)} {m_N\over  8 \pi f^2}\cr
+\qquad\sum_{X=\pi,K,\eta}
{1\over
32\pi^2f^2}\left(\bar\gamma_i^{(X)}-2 \bar\lambda_i^{(X)}
\alpha_i\right)M^2_X\ln M^2_X/\mu^2,
}}
where ${ \a}_i$ are the tree-level predictions derived from
Lagrangian~\cgmag, $\beta_i^{(\pi,K)}$ and $\beta_i^{'(\pi,K)}$ are the
contributions from pion and kaon loops from \figone\ with intermediate
octet and decuplet states respectively, $\bar\gamma_i^{(\pi,K,\eta)}$
are the contributions from  pion, kaon and $\eta$ loops from \figtwo,
and $\bar\lambda_i$ is the wavefunction renormalization contribution.
$\eta$ loops do not contribute to the diagrams of \figone\ since the
$\eta$ is neutral. The scale $\mu$ is an arbitrary renormalization
scale, and is chosen to be $\mu \sim 1$~GeV. The chiral coefficients
$\alpha_i$, $\beta_i^{(X)}$, $\beta_i^{'(X)}$, $\bar\gamma_i^{(X)}$, and
$\bar\lambda_i^{(X)}$ are listed in Appendix~A.  The pion contribution
to the $m_q^{1/2}$ terms is comparable to the kaon contribution, and
cannot be neglected. The pion mass is not large compared with the
$\Delta-N$ mass difference $\delta$, so the full dependence of the
Feynman graph on the ratio $\delta/M_\pi$ must be retained. This
dependence is described by the function $F(M_X,\delta, \mu)$, which is
given explicitly in the appendix. The function $F$ is normalized so
that $F(M_X, 0, \mu)=M_X$. The dependence of $F$ on the renormalization
scale $\mu$ is of the form $\delta \ln \mu^2$. The $\mu$ dependence of the
$\beta$ terms in Eq.~\magb\ is compensated by the $\mu$ dependence of
the $m_q$ independent $\alpha$ terms.\foot{The $\mu$ dependence of the
$\beta$ terms can only be absorbed into the $\alpha$ terms if the
$\delta$ dependence of both the pion and kaon loops is retained, since
$\delta$ is an $SU(3)$ singlet mass parameter. The mass difference
$\delta$ also affects the $m_q\ln m_q$ terms in Eq.~\magb. However, for
these terms, the pion loop contributions are negligible relative to the
kaon and eta contributions, so we have chosen not to compute the
analogous function, and have combined the octet and decuplet
contributions into a single coefficient.} The $\mu$ dependence of the
$m_q\ln m_q$ terms in Eq.~\magb\ is canceled by the $\mu$ dependence of
local counterterms. These counterterms are the most general invariants
that can be constructed out of $B$, $\bar B$, $M$ and $Q F_{\mu\nu}$ which are
linearly independent and preserve parity and time-reversal invariance,
\eqn\cterms{\eqalign{
\CL=&F_{\mu\nu}\Bigl(
c_1 \Tr \bar B M Q\sigma^{\mu\nu} B+c_2
\Tr \bar B Q \sigma^{\mu\nu}B M+c_3 \Tr \bar B\sigma^{\mu\nu}  B M Q
\cr&\qquad+c_4\Tr \bar B M \sigma^{\mu\nu}B Q
+c_5\Tr \bar B \sigma^{\mu\nu}B \Tr M Q\cr
&\qquad+c_6 \Tr M \Tr \bar B
Q\sigma^{\mu\nu}  B+c_7\Tr M \Tr B\sigma^{\mu\nu} B Q\Bigr).
}}
The counterterm with flavor structure $\Tr \bar B Q \ \Tr B M +
\Tr \bar B M \ \Tr B Q$ is a linear combination of the counterterms in
Eq.~\cterms, and the counterterm $i(\Tr \bar B Q\ \Tr B M -  \Tr \bar B
M \ \Tr B Q)$ violates time-reversal invariance.

The theoretical computation can now be compared with experiment. There are
seven octet magnetic moments as well as the
$\Sigma^0\rightarrow\Lambda\gamma$ transition magnetic moment which are
measured experimentally. (The $\Sigma^0$ magnetic moment has not been
measured.) The well-known $SU(3)$ symmetric fit to the data is obtained
by using only the tree level $\alpha_i$ terms in Eq.~\magb, and has an
average deviation between the theoretical and experimental numbers of
0.25 nuclear magnetons. There are eight experimentally measured magnetic
moments, and two parameters $\mu_D$ and $\mu_F$, which leads to the six
linear relations found by Coleman and Glashow \coleglas
\eqn\treereln{\eqalign{
&\mu_{\Sigma^+}=\mu_p\ (2.42\pm0.05=2.79),\cr
&2\mu_\Lambda=\mu_n \ (-1.23\pm0.01=-1.91),\cr
&\mu_{\Xi^0}=\mu_n\ (-1.25\pm0.01=-1.91),\cr
}
\quad\eqalign{
&\mu_{\Sigma^-}+\mu_n=-\mu_p\ (-3.07\pm0.03=-2.79),\cr
&\mu_{\Xi^-}=\mu_{\Sigma^-}\ (-0.65=-1.16\pm0.03),\cr
&2\mu_{\Lambda\Sigma^0}=-\sqrt{3}\mu_n\ (3.22\pm0.16=3.31),\cr
}}
where the relations are written so that all terms in a given relation
have the same sign. The experimental numbers for each relation
are given in parentheses, and have been rounded off to two decimal digits.

The $m_q^{1/2}$ correction is more important than the $m_q\ln m_q$
contribution and the seven $\CO(m_q)$ counterterms. We therefore first
discuss the theoretical predictions including only the $m_q^{1/2}$
corrections, which have the same two unknown parameters $\mu_D$ and
$\mu_F$ as the tree level predictions. There are three relations between
the magnetic moments which are valid to $\CO(m_q^{1/2})$, irrespective
of the baryon-pion axial coupling constants.  These relations, linear
combinations of the six relations of Eq.~\treereln\ valid at tree
level, were noted by Caldi and Pagels \calpag\ and are also valid when
decuplet graphs are included:
\eqn\halfreln{\eqalign{
&\mu_{\Sigma^+}=-2\mu_{\Lambda}-\mu_{\Sigma^-}
\ (2.42\pm0.05=2.39\pm0.03),\cr
&\mu_{\Xi^0}+\mu_{\Xi^-}+\mu_n=2\mu_\Lambda-\mu_p
\ (-3.81\pm0.0=-4.02\pm0.01),\cr
&\mu_\Lambda-\sqrt{3}\mu_{\Lambda\Sigma^0}=\mu_{\Xi^0}+\mu_n
\ (-3.40\pm0.14=-3.16\pm0.01).
}}
The experimental values are shown in parentheses. These relations are in
good agreement with experiment, and work much better than the tree level
relations Eq.~\treereln. The remaining three relations are predictions
for the deviation from any three of the Coleman-Glashow relations, \eg
\eqn\cgdev{\eqalign{
&\mu_p-\mu_{\Sigma^+}= 0.37\pm0.05=\cr
&\quad {m_N\over 8 \pi f^2}
\left[\left(\frac13D^2+2D F-F^2\right)
\left(M_K-M_\pi\right)+\frac5{18}\CC^2 \left( F\left(M_K,
\delta,\mu\right) - F\left(M_\pi,\delta,\mu\right)\right)\right],\cr
&\mu_{\Xi^-}-\mu_{\Sigma^-}= 0.51\pm0.03=\cr
&\quad {m_N\over 8 \pi f^2}
\left[\left(-\frac13D^2+2D F+F^2\right)
\left(M_K-M_\pi\right)+\frac1{18}\CC^2 \left( F\left(M_K,
\delta,\mu\right) - F\left(M_\pi,\delta,\mu\right)\right)\right],\cr
&\mu_{\Xi^0}-\mu_n=0.66\pm0.01=\cr
&\quad {m_N\over 8 \pi f^2}
\left[\left(2 D^2+2 F^2\right)
\left(M_K-M_\pi\right)+\frac1{9}\CC^2 \left( F\left(M_K,
\delta,\mu\right) - F\left(M_\pi,\delta,\mu\right)\right)\right],\cr
}}
The numerical values in Eq.~\cgdev\ are taken from experiment. The
relations Eq.~\cgdev\ are independent of the renormalization scale
$\mu$, which cancels in the difference $F(M_K, \delta, \mu) - F(M_\pi,
\delta, \mu)$. The theoretical predictions using the tree level values
for the axial couplings\foot{ We have used the best fit values $F=0.47$
and $D=0.81$ from \ref\jaffe{R.L. Jaffe and A.V. Manohar,
\np{337}{1990}{509}} and $\CC=-1.53$ from \jmtwo.}\ do not agree at all
with the experimental data. The theoretical values for the relations in
Eq.~\cgdev\ are 1.79, 1.30, and 2.94 nuclear magnetons respectively,
which are much larger than the experimental numbers. A least squares fit
to the eight experimentally measured moments including $m_q^{1/2}$
corrections has an average deviation of 0.8 nuclear magnetons, which is
more than three times larger than the tree level fit.

Naively, one might conclude that the reason for this failure is that the
$K$ mass is too large for chiral perturbation  theory to be valid and
$K$ meson loops should be omitted, or that intermediate decuplet states
should not be included in chiral perturbation theory. Neither of these
two conclusions is substantiated by the data. One can repeat the fit
using Eq.~\magb\ without including the decuplet contributions and the
fit to experiment is still much worse than the tree level fit, the
average deviation from experiment being 0.7 nuclear magnetons. One can
also repeat the fits dropping the $K$ loops completely and retaining
only the pion loops; the average deviation from experiment is 0.7
nuclear magnetons if the decuplet is included, and 0.6 nuclear magnetons
if it is omitted. All of these fits are much worse than the tree level
fit, which has neither pion nor kaon loops. The pion mass is small
enough that chiral perturbation theory in the pion mass is valid, so
there is no theoretical reason why pion loops should be dropped from the
calculation. Pion loop contributions alone are in serious disagreement
with experiment.

We believe that the reason for the disagreement is that the formula Eq.~\magb\
overestimates the size of kaon loops. The suppression of kaon loops in
chiral perturbation theory has been discussed before by Gasser and
Leutwyler \gasleut. They have suggested a method for evaluating loop
graphs that suppresses kaon loops, which they call improved chiral
perturbation theory (ICPT). There is some empirical evidence that kaon
loops are suppressed. Loop corrections to the baryon masses and hyperon
non-leptonic decays \j\ work well if the one-loop corrected values for
the axial vector coupling constants \jm\ (which are smaller than the
tree level values) are used.  A similar result is found for the nucleon
polarizability \bsone. The one-loop couplings $D$, $F$, and $\CC$ are
proportional to, but smaller than, their tree level values. Another well
known effect which also suppresses kaon loops is that $f_K=1.2 f_\pi$.
We have therefore computed the results using the central values for the
axial couplings extracted at one loop $D=0.61$, $F=0.4$ and $\CC=-1.2$
given in \jmtwo\ and \bss, and $f_K=1.2\,f_\pi$. The deviations from the
Coleman-Glashow relations are now 0.59, 0.50 and 1.0 nuclear magnetons,
and are much closer to the experimental values. One can also do a least
squares fit to the data, which has the same two unknown parameters as
the tree level fit. The average deviation is 0.11 nuclear magnetons,
about half that of the tree level fit.

Finally, one can include the $m_q\ln m_q$ contributions to the baryon
magnetic moments. Since $\ln M_K^2/\mu^2$ is not very big, the $m_q \ln
m_q$ terms are not expected to be significantly enhanced relative to the
counterterms in Eq.~\cterms, so we will include the counterterms in the
theoretical predictions.  There are seven unknown
counterterm coefficients $c_1$--$c_7$, as well as the two tree level
parameters $\mu_D$ and $\mu_F$. The $c_6$ and $c_7$ counterterms have
the same vector $SU(3)$ structure as the lowest order terms, so there
are effectively seven independent operators, and there exists one linear
relation amongst the baryon magnetic moments which is valid including
all terms of order $m_q^{1/2}$, $m_q\ln m_q$ and $m_q$,
\eqn\final{
6\mu_\Lambda+\mu_{\Sigma^-}-4\sqrt{3}\mu_{\Lambda\Sigma^0} = 4\mu_n -
\mu_{\Sigma^+} + 4\mu_{\Xi^0}\ (-15.99\pm0.56 =-15.07\pm0.08) } which
agrees well with experiment. In addition to the relations of
Eqs.~\treereln--\final, there is also the $SU(2)$ relation
\eqn\sutwo{
\mu_{\Sigma^+} + \mu_{\Sigma^-} = 2 \mu_{\Sigma^0},
} which cannot be tested because the $\Sigma^0$ magnetic moment has not
been measured.

\bigskip

In conclusion, we have calculated the nonanalytic contributions
proportional to $m_q^{1/2}$ and $m_q \ln m_q$ to the baryon magnetic
moments in chiral perturbation theory.  We find that the one-loop
fit to the data with the $m_q^{1/2}$ corrections is much worse than the
tree-level result if the tree level axial couplings are used, but is
better than the tree level result if the one loop axial couplings are
used. This suggests that kaon loop graphs are overestimated in chiral
perturbation theory when the tree-level couplings are used at the
vertices. Unlike the cases of the baryon axial couplings \jm\ and masses
\j, including virtual baryon  decuplet states does not give appreciably
better agreement with the data. Relations amongst the magnetic moments
which are independent of the axial couplings are found to be in good
agreement with experiment.

\bigskip\bigskip
\centerline{\bf Acknowledgements}\nobreak
This work was supported in part by the Department of Energy under grant
number DOE-FG03-90ER40546.  AVM was also supported by a National Science
Foundation Presidential Young Investigator award PHY-8958081.  MJS
acknowledges support of a Superconducting Supercollider National
Fellowship from the Texas National Research Laboratory Commission under
grant FCFY9219.

\vfill\break\eject

\appendix{A}{Chiral Coefficients}
\noindent The tree level coefficients are:
\eqn\amag{\eqalign{
&\a_p = \frac 1 3 \mu_D + \mu_F, \cr &\a_n = -\frac 2 3 \mu_D, \cr
&\a_{\Lambda} = -\frac 1 3 \mu_D, \cr}
\quad \eqalign{
&\a_{\Sigma^+} = \frac 1 3 \mu_D + \mu_F, \cr &\a_{\Sigma^0} = \frac 1 3
\mu_D, \cr &\a_{\Sigma^-} = \frac 1 3 \mu_D - \mu_F, \cr}
\quad \eqalign{
&\a_{\Xi^0} = -\frac 2 3 \mu_D, \cr &\a_{\Xi^-} = \frac 1 3 \mu_D -
\mu_F, \cr &\a_{\Lambda \Sigma^0} = \frac 1 {\sqrt{3}} \mu_D .\cr }} The
one-loop coefficients $\b_i$ and $\b^\prime_i$
from the graphs in \figone$(a)$ and $(b)$ are:
\eqn\bmagp{\eqalign{
&\bpi_p = -(D+F)^2, \cr &\bpi_n = (D+F)^2,
\cr &\bpi_{\Lambda} = 0, \cr}
\quad \eqalign{
&\bpi_{\Sigma^+} = -\frac 2 3 D^2 - 2 F^2, \cr
&\bpi_{\Sigma^0} = 0, \cr &\bpi_{\Sigma^-} = \frac 2 3 D^2 + 2 F^2, \cr}
\quad \eqalign{
&\bpi_{\Xi^0} = -(D-F)^2, \cr &\bpi_{\Xi^-} = (D-F)^2, \cr
&\bpi_{\Lambda{\Sigma^0}} = - \frac 4{\sqrt 3}
DF,\cr }}
\eqn\bpmagp{\eqalign{
&\beta^{'(\pi)}_p =-\frac 2 9\c^2, \cr
&\beta^{'(\pi)}_n = \frac 2 9 \c^2,
\cr &\beta^{'(\pi)}_{\Lambda} = 0, \cr}
\qquad \eqalign{
&\beta^{'(\pi)}_{\Sigma^+} =\frac 1{18} \c^2, \cr
&\beta^{'(\pi)}_{\Sigma^0} = 0, \cr &\beta^{'(\pi)}_{\Sigma^-} =-
\frac 1{18} \c^2, \cr}
\qquad \eqalign{
&\beta^{'(\pi)}_{\Xi^0} =\frac 1 9 \c^2, \cr &\beta^{'(\pi)}_{\Xi^-} = -
\frac 19 \c^2, \cr &\beta^{'(\pi)}_{\Lambda{\Sigma^0}} =
-\frac1{3\sqrt3}\c^2,\cr }}
for the pion loops, and
\eqn\bmagk{\eqalign{
&\bk_p = -\frac 2 3 D^2 - 2 F^2, \cr &\bk_n =
-(D-F)^2, \cr &\bk_{\Lambda} = 2 D F, \cr}
\quad \eqalign{
&\bk_{\Sigma^+} = -(D+F)^2, \cr &\bk_{\Sigma^0} = -2 D
F, \cr &\bk_{\Sigma^-} = (D-F)^2, \cr}
\quad \eqalign{
&\bk_{\Xi^0} = (D+F)^2, \cr &\bk_{\Xi^-} = \frac 2 3 D^2
+ 2 F^2, \cr &\bk_{\Lambda\Sigma^0} = - \frac 2{\sqrt
3} DF,\cr }}
\eqn\bpmagk{\eqalign{
&\beta^{'(K)}_p =\frac 1 {18} \c^2, \cr &\beta^{'(K)}_n =
\frac 1 9 \c^2, \cr &\beta^{'(K)}_{\Lambda} =\frac 1 6 \c^2, \cr}
\qquad \eqalign{
&\beta^{'(K)}_{\Sigma^+} = - \frac 2 9 \c^2, \cr &\beta^{'(K)}_{\Sigma^0} = -
\frac 1 6 \c^2, \cr
&\beta^{'(K)}_{\Sigma^-} =- \frac 1 9 \c^2, \cr}
\qquad \eqalign{
&\beta^{'(K)}_{\Xi^0} = \frac 2 9 \c^2, \cr &\beta^{'(K)}_{\Xi^-} = - \frac 1
{18} \c^2, \cr &\beta^{'(K)}_{\Lambda\Sigma^0} = -\frac 1{6\sqrt3}\c^2,\cr }}
for the graphs with kaon loops.  The function $F(M,\delta,\mu)$ is
\eqn\fndef{
\pi F\left(M,\delta,\mu\right)=\cases{-\delta\ln M^2/\mu^2 + 2
\sqrt{M^2-\delta^2}\left[{\pi\over2}-
\tan^{-1}{\delta\over\sqrt{M^2-\delta^2}}\right]&$\delta \le M$,\cr
\noalign{\medskip}
-\delta\ln M^2/\mu^2 + \sqrt{\delta^2-M^2}\ln{\delta-\sqrt{\delta^2-M^2}
\over\delta+\sqrt{\delta^2-M^2}}&$\delta > M$.\cr}
}
The
wavefunction renormalization coefficients are $SU(2)$ invariant, and
the values for the different $SU(2)$ multiplets are:
\eqn\wave{\eqalign{
&\bar\lambda_N^{(\pi)}=\frac 9 4 (D+F)^2+2\c^2\cr
&\bar\lambda_\Lambda{(\pi)}= 3
D^2+\frac32\c^2\cr
\noalign{\medskip}
&\bar\lambda_N^{(K)}=\frac 5 2 D^2 - 3 D F + \frac 9 2 F^2+\frac12\c^2\cr
&\bar\lambda_\Lambda^{(K)}= D^2+9 F^2+\c^2\cr
\noalign{\medskip}
&\bar\lambda_N^{(\eta)}=\frac 1 4 (D-3F)^2 \cr
&\bar\lambda_\Lambda^{(\eta)}=D^2\cr}
\quad\eqalign{
&\bar\lambda_\Sigma^{(\pi)}=D^2 + 6 F^2+\frac13\c^2\cr
&\bar\lambda_\Xi^{(\pi)}=\frac 9 4 (D-F)^2+\frac12\c^2\cr
\noalign{\medskip}
&\bar\lambda_\Sigma^{(K)}=3 D^2 +3 F^2+\frac53\c^2\cr
&\bar\lambda_\Xi^{(K)}=\frac 5 2 D^2 + 3 D F + \frac 9 2 F^2+\frac32\c^2\cr
\noalign{\medskip}
&\bar\lambda_\Sigma^{(\eta)}=D^2+\frac12\c^2\cr
&\bar\lambda_\Xi^{(\eta)}=\frac
1 4 (D+3F)^2+\frac12\c^2 \cr }} The wavefunction renormalization
coefficients for the transition magnetic moment are
$\bar\lambda_{\Lambda\Sigma^0}^{(X)}=\half(\bar\lambda_{\Lambda}^{(X)}+
\bar\lambda_{\Sigma}^{(X)})$, for $X=\pi,\ K,\ \eta$.
The coefficients $\bar\gamma_i$ evaluated from the graphs in \figtwo\ are:
\eqn\deltap{\eqalign{
&\bar\gamma^{(\pi)}_p =
-\mu_D-\mu_F+\frac12(D+F)^2(\mu_D-\mu_F)-\frac{32}{27}\c^2\mu_C+\frac{8}9
\c(D+F)\mu_T, \cr
&\bar\gamma^{(\pi)}_n = -(D+F)^2\mu_F+\frac8{27}\c^2\mu_C-\frac{8}9\c(D+F)
\mu_T, \cr
&\bar\gamma^{(\pi)}_{\Lambda} = -\frac13\mu_D-\frac 23 D^2\mu_D-
\frac23\c D\mu_T, \cr
&\bar\gamma^{(\pi)}_{\Sigma^+} = -\mu_D - \mu_F+
\frac 29(D^2 + 6 D F - 6 F^2)\mu_D - 2 F^2\mu_F-\frac2{27}\c^2\mu_C
+\frac29\c(D+3F)\mu_T, \cr
&\bar\gamma^{(\pi)}_{\Sigma^0} = -\mu_D+\frac
29(D^2-6F^2)
\mu_D+\frac{4}{9}\c F\mu_T, \cr
&\bar\gamma^{(\pi)}_{\Sigma^-} = -\mu_D+\mu_F+\frac 29 (D^2 - 6 D F - 6
F^2)\mu_D + 2 F^2\mu_F+\frac2{27}\c^2\mu_C+\frac29\c(F-D)\mu_T,\cr
&\bar\gamma^{(\pi)}_{\Xi^0} =(D-F)^2 \mu_F+\frac4{27}\c^2\mu_C+
\frac29\c(F-D)\mu_T,\cr
&\bar\gamma^{(\pi)}_{\Xi^-} = -\mu_D+\mu_F+\frac 1 2(D-F)^2 (\mu_D +
\mu_F)+\frac2{27}\c^2\mu_C
+\frac{4}9\c(F-D)\mu_T, \cr &\bar\gamma^{(\pi)}_{\Lambda \Sigma^0} =
-\frac1{\sqrt{3}}\mu_D+\frac 2 {3\sqrt{3}}D(6 F \mu_F-D
\mu_D)-\frac4{9\sqrt{3}} \c^2\mu_C+\frac1{9\sqrt{3}}\c(D+6F) \mu_T,
}} for the pion loops,
\eqn\deltak{\eqalign{
&\bar\gamma^{(K)}_p = -\mu_D-\mu_F+ (-\frac19D^2+ 2DF-F^2) \mu_D-
(D-F)^2\mu_F-\frac4{27}\c^2\mu_C
\cr &\qquad\qquad +\frac29\c(3D-F)\mu_T, \cr
&\bar\gamma^{(K)}_n = (-\frac79D^2+ \frac23DF+F^2)\mu_D +
(D-F)^2\mu_F+\frac{4}{27}\c^2\mu_C -\frac{4}9\c F\mu_T, \cr
&\bar\gamma^{(K)}_{\Lambda} = -\frac13\mu_D+(\frac19D^2+ F^2)
\mu_D-2DF \mu_F+\frac29\c^2\mu_C
+\frac29\c(D-3F)\mu_T, \cr
&\bar\gamma^{(K)}_{\Sigma^+} =-\mu_D - \mu_F+
(\frac13D^2+ 2DF+\frac13F^2)\mu_D - (D-F)^2\mu_F-\frac{28}{27}
\c^2\mu_C+\frac{8}9\c D\mu_T, \cr &\bar\gamma^{(K)}_{\Sigma^0} =
-\mu_D+(\frac13D^2+\frac13F^2) \mu_D+ 2 D F\mu_F-\frac29\c^2\mu_C
+\frac{2}9\c(D+F)\mu_T,
\cr &\bar\gamma^{(K)}_{\Sigma^-} = -\mu_D+\mu_F+(\frac13D^2-
2DF+\frac13F^2)\mu_D + (D+F)^2\mu_F+
\frac{16}{27}\c^2\mu_C\cr&\qquad\qquad +\frac{4}9\c(F-D)\mu_T, \cr
&\bar\gamma^{(K)}_{\Xi^0} =(-\frac79D^2- \frac23DF+F^2)\mu_D -
(D+F)^2\mu_F+\frac8{27}\c^2\mu_C -\frac{4}9\c(D+2F)\mu_T, \cr
&\bar\gamma^{(K)}_{\Xi^-} =-\mu_D+\mu_F+ (-\frac19D^2- 2DF-F^2) \mu_D+
(D+F)^2\mu_F+\frac{16}{27}\c^2\mu_C+\cr&\qquad\qquad
\frac29\c(F-D)\mu_T, \cr
&\bar\gamma^{(K)}_{\Lambda \Sigma^0} =
-\frac1{\sqrt{3}}\mu_D+\frac1{\sqrt{3}}(D^2-3F^2) \mu_D+
\frac2{\sqrt{3}}DF\mu_F
-\frac2{9\sqrt{3}}\c^2\mu_C +\frac{4}{9\sqrt{3}}\c(2D+3F)\mu_T, }} for
the kaon loops, and
\eqn\deltae{\eqalign{
&\bar\gamma^{(\eta)}_p = -\frac1{18}(D-3F)^2(\mu_D+3\mu_F), \cr
&\bar\gamma^{(\eta)}_n = \frac19(D-3F)^2\mu_D, \cr
&\bar\gamma^{(\eta)}_{\Lambda} = \frac29D^2\mu_D, \cr
&\bar\gamma^{(\eta)}_{\Sigma^+} =
-\frac29D^2(\mu_D+3\mu_F)-\frac29\c^2\mu_C+
\frac{4}9\c D\mu_T, \cr
&\bar\gamma^{(\eta)}_{\Sigma^0} = -\frac29D^2\mu_D+\frac29\c D\mu_T, \cr
&\bar\gamma^{(\eta)}_{\Sigma^-} =
-\frac29D^2(\mu_D-3\mu_F)+\frac29\c^2\mu_C, \cr
&\bar\gamma^{(\eta)}_{\Xi^0}
=\frac19(D+3F)^2\mu_D-\frac29\c(D+3F)\mu_T, \cr
&\bar\gamma^{(\eta)}_{\Xi^-}
= - \frac1{18}(D+3F)^2(\mu_D-3\mu_F)+\frac29\c^2\mu_C,
\cr
&\bar\gamma^{(\eta)}_{\Lambda \Sigma^0} =
\frac2{3\sqrt{3}}D^2\mu_D+\frac1{3\sqrt{3}}\c D\mu_T,
}} for the $\eta$ loops.  All the loop coefficients include
contributions from intermediate octet and decuplet states.  The values
of the coefficients neglecting all decuplet contributions are
obtained trivially by setting $\CC=0$.

\listrefs
\listfigs
\insertfig{Figure 1}{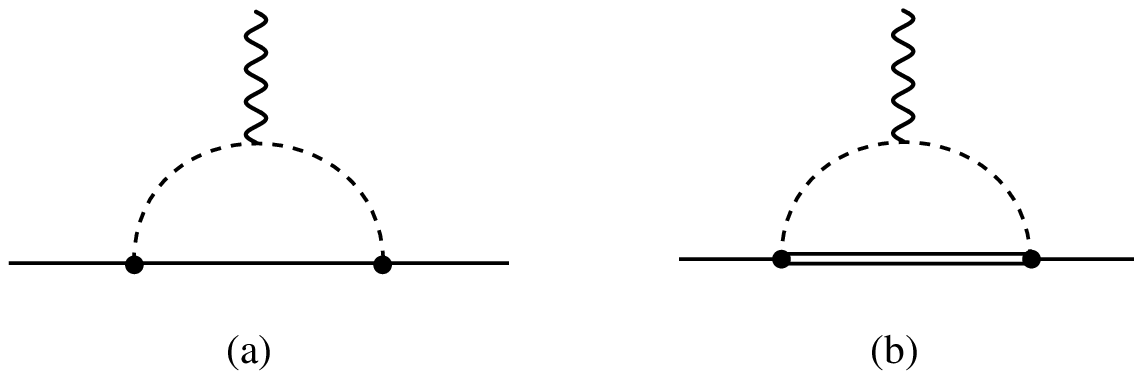}
\insertfig{Figure 2}{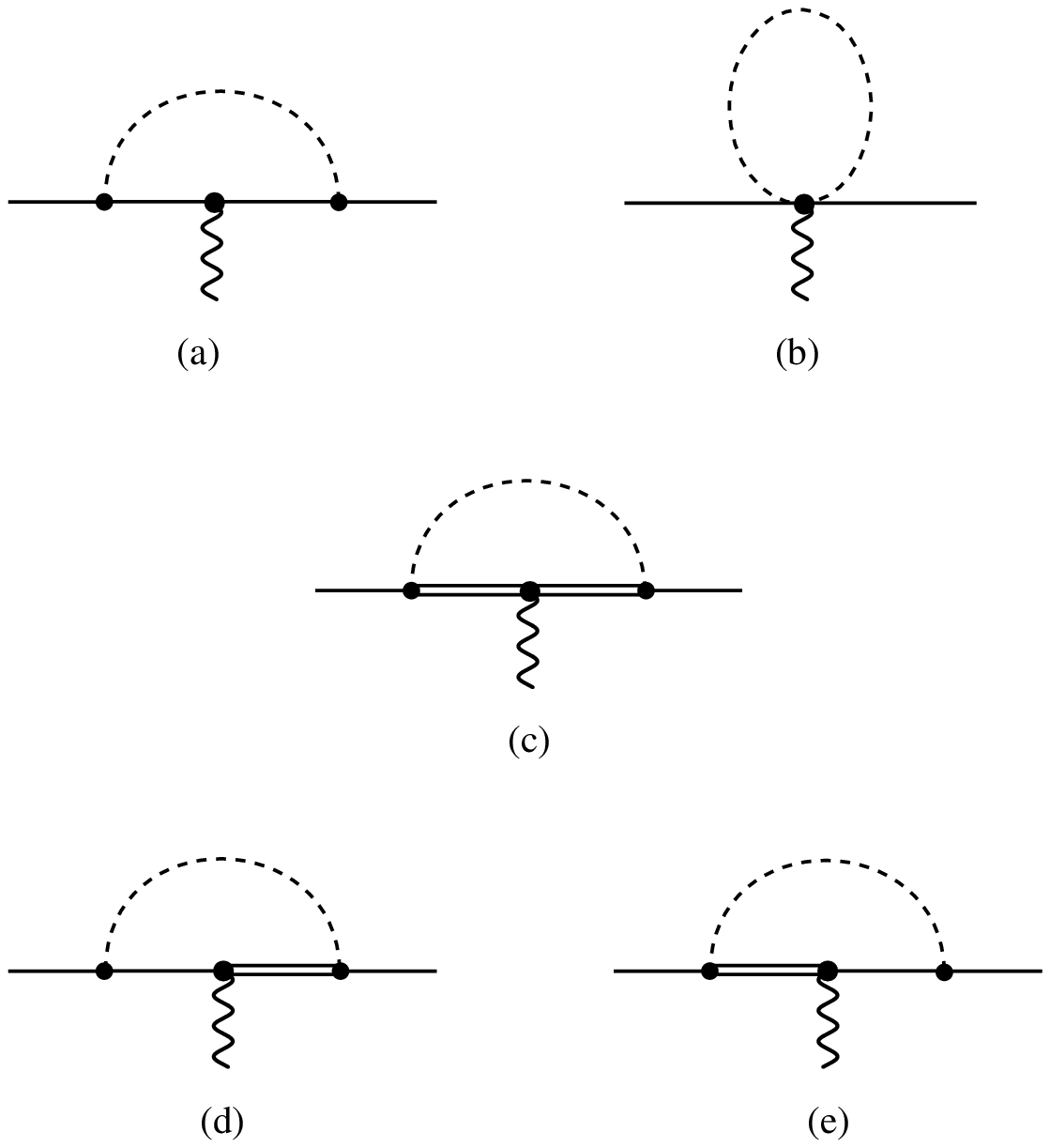}
\bye